\documentclass[%
 reprint,
 showpacs, showkeys,
 amsmath,amssymb,
 aps,prc,
]{revtex4-1}
\usepackage{amsmath}
\usepackage{cases}
\usepackage{amssymb}
\usepackage{CJK}
\usepackage{graphicx}
\usepackage{dcolumn}
\usepackage{bm}
\usepackage{color}
\usepackage[pagewise]{lineno}

\begin{document}
\begin{CJK*}{GBK}{} 

\preprint{APS/123-QED}

\title{Neutron-skin effects in isobaric yield ratio for mirror nuclei in statistical abrasion-ablation model}

\author{Chun-Wang Ma$^{1}$}\thanks{Email: machunwang@126.com}
\author{Hui-Ling Wei$^{1}$}
\author{Yu-Gang Ma$^{2}$}\thanks{Email: ygma@sinap.ac.cn}

\affiliation{$^{1}$ Institute of Particle and Nuclear Physics, Henan Normal University, \textit{Xinxiang 453007}, China\\
$^{2}$ Shanghai Institute of Applied Physics, Chinese Academy of Sciences, \textit{Shanghai 201800}, China
}




\date{\today}

\begin{abstract}
\begin{description}
\item[Background] The isobaric yield ratio for mirror nuclei [IYR(m)] in heavy-ion collisions,
    which is assumed to depend linearly on $x = 2(Z-1)/A^{1/3}$ of a fragment, is applied to study some
    coefficients of the energy terms in the binding energy, as well as the difference between the chemical potentials
    of a neutron and proton. It is found that the IYR(m) has a systematic dependence on the reaction,
    which has been explained as the volume and/or the isospin effects in previous studies. However,
    neither the volume nor the isospin effects can fully interpret the data.
\item[Purpose] We suppose that the IYR(m) depends on the neutron-skin thickness ($\delta_{np}$) of
    the projectile, and check the idea of whether the neutron-skin thickness effects
    can fully explain the systematic dependence of the IYR(m).
\item[Methods] A modified statistical abrasion-ablation model is used to calculate the reactions induced by projectiles
    of three series: (1) the calcium isotopes from $^{36}$Ca to $^{56}$Ca as projectiles with different
    limitations on the impact parameters ($b_{\mbox{max}}$) to show the volume effects according to $b_{\mbox{max}}$;
    (2) the $A = 45$ isobars as the projectiles having different isospins and $\delta_{np}$;
    and (3) projectiles having similar $\delta_{np}$ to show whether the IYR(m) depends on the volume or
    the isospin of the projectile.
\item[Results] The IYR(m) shows a distribution of a linear part in the small-$x$ fragments, and a nonlinear part
    in the large-$x$ fragments. The linear part of IYR(m) is fitted. (1) In the calcium isotopic reactions, the
    IYR(m) depends on the isospin or the volume of the projectile, but $\delta_{np}$ greatly influences the
    nonlinear part of the IYR(m). The IYR(m) does not depend on the colliding source in reactions of small $b_{\mbox{max}}$
    for the nonneutron-rich projectiles, and does not depend on the collision sources in reactions by the neutron-rich projectiles;
    (2) In reactions of the $A = 45$ isobars, though IYR(m) depends on the isospin of projectile,  IYR(m) shows small
    dependence on isospin if $\delta_{np} > 0$; (3) In the reactions of projectiles having similar $\delta_{np}$, the
    IYR(m) in the small mass fragments show no dependence on the volume and the isospin of the projectile when the
    mass of the projectile is relatively large. Specially, the dependence of IYR(m) on the mass of the isospin of the projectile
    vanishes when $\delta_{np}\sim0.02$fm.
\item[Conclusions] The linear and nonlinear parts of the IYR(m) are governed by the core and the surface (skin)
    of the projectile, respectively. The neutron-skin effects can well explain the systematic dependence of the IYR(m).
\end{description}
\end{abstract}

\maketitle
\end{CJK*}


\section{introduction}

In studying the nuclear symmetry energy, the isobaric methods have attracted much
attention recently. The energy terms contributing to the binding energy of a nucleus or fragment, which only depend on the mass number, cancel out in the difference between the binding energies of isobars. This makes the isobaric methods possible to
study the retained terms in the mass formula. For example, the symmetry energy of neutron-rich nucleus
is studied via the difference between binding energies of isobars \cite{isobsymmNPA07,MeiHJPG10,MA12CPL09IYRAsbsbv}. In models based on free
energy, the yield of a fragment is determined by its free energy, the properties
of the colliding source, and the temperature \cite{ModelFisher3,Huang-PRC11-freeenergy,PMar12PRCIsob-sym-isos}.
The symmetry energy of fragment at finite temperature in heavy-ion
collisions (HICs), which has a finite temperature, is also studied using the isobaric yield ratio (IYR) methods.
After the work using the IYR to study the ratio of
the symmetry energy coefficient to the temperature ($a_{\mbox{sym}}/T$) of a fragment \cite{Huang10}, the results using the IYR methods are also discussed using the statistical multifragmentation model \cite{PMar-IYR-sym13PRC},
the canonical and the grand canonical ensembles methods \cite{Mallik13-sym-IYR,Souza12finite},
and free-energy-based models \cite{Huang-PRC11-freeenergy,PMar12PRCIsob-sym-isos}. Moreover, the
IYR methods are also used to study the $a_{\mbox{sym}}/T$ of neutron-rich fragments
\cite{MaCW11PRC06IYR,MaCW12CPL06,MaCW12EPJA,MaCW13CPC}, the formation time of
fragments \cite{Huang10PRCTf,Huang-mscaling}, the difference between the chemical
potentials of a neutron and proton \cite{MaCW13isoSB}, and the temperature of the
heavy fragments \cite{MaCW12PRCT}.

In particular, the IYR for mirror nuclei [IYR(m)], the volume-, surface-, and symmetry-energy
terms contributing to the free energy cancel out. The IYR(m) can be written as follows \cite{Huang10,MaCW11PRC06IYR,MaCW13IYRmFN}:
\begin{equation}\label{lnRmirror}
\mbox{IYR(m)}=\mbox{ln}(Y_2/Y_1)= (\Delta\mu+a_c\cdot x)/T,
\end{equation}
with $Y_2$ and $Y_1$ being the yields of the $I = 1$ and $-1$ ($I = N - Z$
is the neutron excess) fragments, respectively; $\Delta\mu = \mu_n-\mu_p$,
$\mu_n$ and $\mu_p$ being the chemical potentials of the neutron and proton,
respectively. $a_c$ is the Coulomb-energy coefficient and $T$ is temperature;
IYR(m) depends linearly on $x$, with $x = 2(z+1/2)/A^{1/3}$ ($x$ is the charge
number of the $I = -1$ fragment) as in Ref. \cite{Souza12finite} and $x = 2(z-1)/A^{1/3}$
in Ref. \cite{Huang10,MaCW11PRC06IYR,MaCW12CPL06,MaCW12EPJA,MaCW13CPC} since
different form of Coulomb energy are adopted, but it is confirmed that the two
choices of $x$ introduce a very small difference. The $a_c\cdot x$ can be seen as
the residue Coulomb interaction (RCI) between the related isobars \cite{Souza12finite}.

Based on  Eq. (\ref{lnRmirror}), the values of $a_c/T$ and $\Delta\mu/T$ can be
obtained from the IYR(m). A linear correlation between the IYR(m) and $(Z/A)_{\mbox{sys}}$
of the reaction system is used to determine $a_c/T$ and $\Delta\mu/T$ \cite{Huang10}.
Marini \textit{et al.} provided a method to figure out the RCI by fitting the
difference between IYRs \cite{PMar12PRCIsob-sym-isos}. 
It has been concluded that the IYR(m) depends on the volume of the reaction
systems using the standard grand-canonical and canonical statistical ensembles
(SGC/CSE) theories, which prevents the IYR method from obtaining the actual
values from fitting nuclear collision data \cite{Souza12finite}. But conclusions
disagree with the SGC/CSE theories proposed in a modified statistical
abrasion-ablation (SAA) model by considering the density difference in the
projectile, i.e., the IYR(m) depends on the isospin of the projectile \cite{MaCW13IYRmFN}.
At the same time, the SGC/CSE is also shown in part disagreement with the experimental
results \cite{MaCW13IYRmFN}. Due to the contradiction in the SGC/CSE, experimental
and SAA results, it is meaningful to investigate the system dependence of IYR(m) in
different reactions.

Believing the importance of density distribution in determining the yields of fragments
and the resultant parameters, such as $\Delta\mu$ and $a_c$, in this article, we focus
on the investigation of the neutron-skin effects in IYR(m). The SAA model will be used
because it can well reproduce the yield of the fragment \cite{AMDYD08PRC,AA-Yield-acc,MaCW09CPB,WHL10},
though the SAA model does not include the complex evolution process like the antisymmetrized
molecular dynamics models\cite{AMDYD08PRC}. The article is organized as follows. The
SAA model is briefly introduced in Sec. \ref{model}. The results and discussion are
given in Sec. \ref{results}, and a summary is presented in Sec. \ref{summary}.

\section{model description}
\label{model}
In brief, the SAA model is a two-stages model to that predicts the yield of fragments in
reactions above the Fermi energy. The first stage describes the colliding, in which
the abraded nucleons and the yield of the hot prefragment are determined. The second
stage is the evaporation after which the final fragments are formed. It can well
reproduce the yield of fragments and is used in studying the isospin phenomena
in HICs \cite{MaCW09CPB,WHL10,SAABrohm94,SAAGaim91,MaCW09PRC,FangPRC00}.

In the colliding stage, the nuclei are described to be composed of parallel tubes orienting along the beam direction.
The SAA takes independent nucleon-nucleon collisions as the  participants in the overlap zone of the projectile and target
nuclei and determines the distributions of abraded neutrons and protons. For an infinitesimal tube in the projectile,
the transmission probabilities for neutrons (protons) at a given impact parameter $\vec b$ are given by,
\begin{equation}\label{trans}
t_k(\vec s-\vec b)=\mbox{exp}\{-[\rho{_n^T}(\vec s-\vec
b)\sigma_{nk}+\rho{_n^P}(\vec s-\vec b)\sigma_{pk}]\},
\end{equation}
where $\rho^T$ is the nuclear-density distribution of the target integrated along the beam direction, the vectors $\vec s$
and $\vec b$ are defined in the plane perpendicular to the beam, and $\sigma_{k'k}$ is the free nucleon-nucleon reaction cross
section. At a given $\vec b$, the average absorbed mass in the limit of infinitesimal tubes is,
\begin{eqnarray}
<\Delta A(b)>=\int d^{2}s \rho_{n}^{T}(\vec s)[1-t_n(\vec s-\vec b)] \nonumber\\
+\int d^{2}s \rho_{p}^{P}(\vec s)[1-t_p(\vec s-\vec b)] .
\end{eqnarray}
The $\rho_n$ and $\rho_p$ distributions are assumed to be the Fermi-type, 
\begin{equation}\label{Fermi}
\rho_i(r)=\frac{\rho_i^0}{1+\mbox{exp}(\frac{r-C_i}{t_if_i/4.4})},
~~i=n,p ,
\end{equation}
where $\rho_i^0$ is the normalization constant, $t_i$ is the diffuseness parameter, and $C_i$ is the
radius at half density of the neutron or proton density distribution. $t_i$ and $C_i$ can be adjusted by $f_i$
to change the neutron skins thickness ($\delta_{np}$) of a nucleus
\cite{MaCW10PRC,MaCW08CPB,DQF10PRC-nskin-SAA,MaCW11CPC,PST12-isospin-saa,MACW10JPG}.
$\delta_{np}$ of a nucleus is defined as the difference between the root-mean-square radii of the
neutrons' and protons' density distributions ($\delta_{np}= <r_n^2>^{1/2}-<r_p^2>^{1/2}$). In this
work we use the default values of the parameters in Eq. (\ref{Fermi}), and $\delta_{np}$ of the nucleus is not
strictly set to the predicted experimental or theoretical one due to the fact that the measurement
of $\delta_{np}$ itself is still an open question.

The cross section for a specific isotope (prefragment) can be calculated from
\begin{equation}\label{yieldisotope}
\sigma(\Delta N, \Delta Z)=\int d^2bP(\Delta N, b)P(\Delta Z,b),
\end{equation}
where $P(\Delta N, \mathit{b})$ and $P(\Delta Z, \mathit{b})$ are the probability distributions for the abraded neutrons and
protons at a given impact parameter $\mathit{b}$, respectively.

The second stage of the reaction in SAA is the evaporation of the excited prefragment \cite{SAAGaim91},
which is described by a conventional statistical model under the assumption of thermal equilibrium.
The excitation energy of the projectile spectator is estimated by a simple relation of $E^{*}=13.3<A(b)>$MeV,
where 13.3 is the mean excitation energy due to an abraded nucleon from the initial projectile \cite{SAAGaim91}.
After the evaporation, the isotopic yield (final fragment) comparable to the experimental result can be obtained.
It is shown that the deexcitation or decay descriptions (GEMINI, SMM, SIMON, etc.) greatly influence the yield of the
fragment, and the parameters based on the yield  \cite{Huang10,MaCW13CPC,PMar-IYR-sym13PRC,AMDYD08PRC}. Specially,
the prefragment and final fragment in the $^{60}$Ni + $^{12}$C reaction, the resultant IYR and $a_{\mbox{sym}}/T$ of the
prefragment and the final fragment are analyzed, which show that the deexciation process affects the results greatly \cite{MaCW13CPC}.
Since the decay mode in SAA can well reproduce the yield of the final fragment, which will be used in the analysis, the
effect of different decay mechanism will not be discussed in this work.

\section{results and discussion}
\label{results}

The isospin [$I' = (N-Z)/A$], volume (or mass) and neutron-skin effects in IYR(m) will be studied
in the SAA model. The reactions induced by projectiles of three series will be calculated: (1) the
isotopic projectiles to study the isospin and volume effects in IYR(m), at the same time, different
limitations on the maximum of impact parameters will be used to study the volume dependence of IYR(m);
(2) the isobaric projectiles to study the isospin and neutron-skin effects in IYR(m); and (3)
projectiles having similar neutron-skin thickness to study the volume and isospin effects in
IYR(m).

\begin{figure}[htbp]
\includegraphics
[width=8.6cm]{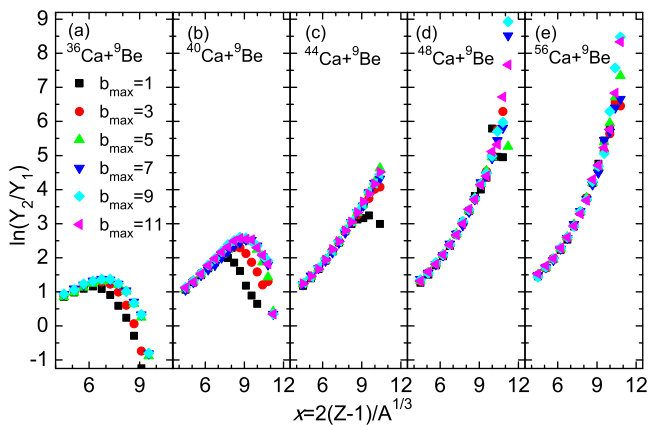}
\caption{\label{Ca3656bmax} (Color online)
The IYR(m) in the 140$A$ MeV $^{36, 40, 44, 48, 56}$Ca + $^9$Be reactions with different
limitations of maximum impact parameters ($b_{\mbox{max}}$) calculated using the SAA model.
$B_{\mbox{max}}$ is changed from 1 to 11fm in the step of 2fm.
}
\end{figure}

First, the 140$A$ MeV $^{36, 40, 44, 48, 56}$Ca + $^9$Be reactions are calculated to study the isospin
dependence of the IYR(m) in the isotopic projectiles. For the projectile from $^{36}$Ca to $^{56}$Ca,
$\delta_{np}$ are --0.117, --0.05, 0.005, 0.053, and 0.129fm, respectively, and $I'$ changes from --0.1
to 0.4. Considering the multiple sources collisions which have different volumes according to the impact
parameters, the volume dependence of the IYR(m) should be manifested. The limitations on the maximum of
impact parameter ($b_{\mbox{max}}$) is varied from 1fm to 11fm in the step of 2fm in
the calculation. In Fig. \ref{Ca3656bmax}, the IYR(m) in these reactions using different $b_{\mbox{max}}$ are
plotted. The IYR(m) shows the distribution as a linear increasing part plus a nonlinear part as $x$
increases. Though the IYR(m) in the $^{36, 40}$Ca reactions are easily influenced by $b_{\mbox{max}}$, the linear
part of IYR(m) changes very little. The IYR(m) in the neutron-rich $^{48, 56}$Ca reactions are scarcely
affected by $b_{\mbox{max}}$. The IYR(m) in the $^{44}$Ca reactions of $b_{\mbox{max}} = 1$fm only shows very little difference
to those of the other $b_{\mbox{max}}$. In the reactions of neutron-rich projectiles, the volume dependence
of the IYR(m) disappears when $b_{\mbox{max}}$ changes.
\begin{figure}[htbp]
\includegraphics
[width=8.6cm]{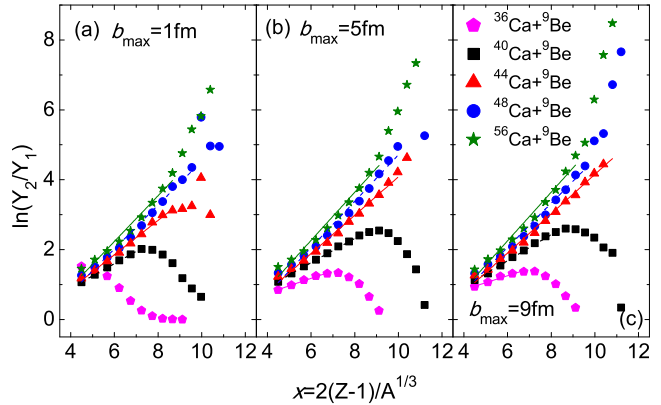}
\caption{\label{Ca3656SPbmax} (Color online)
The IYR(m) in the calculated 140$A$ MeV $^{36, 40, 44, 48, 56}$Ca + $^9$Be reactions
with limitations of $b_{\mbox{max}} =$ 1, 5, and 9,
respectively. The lines are the linear fitting results of IYR(m).
}
\end{figure}

In Fig. \ref{Ca3656SPbmax}, the results in Fig. \ref{Ca3656bmax} are re-plotted according
to $b_{\mbox{max}}$. The IYR(m) changes according to the colliding volumes in the isotopic reactions.
It is clearly shown that, in the large-$x$ fragments, the trend of the nonlinear part of IYR(m)
changes from decreasing to increasing as $x$ increases. It can not be definitely explained
whether the IYR(m) fully depends on the isospin or the volume of the projectile as has been
discussed in Ref. \cite{MaCW13IYRmFN}. Generally, the structure of a nucleus can be considered
as a core plus a skirt region: in the core region, the density changes little, while in the
surface the density decreases quickly. According to Eq. (\ref{yieldisotope}), at a specific incident
energy, the isotopic yield is mainly determined by the $\rho_n$ and $\rho_p$ distributions at
a specific incident energy. In the surface region, the quick change of density has a great
influence on the yield of the fragment in the (semi-)peripheral reactions. In the neutron-rich nucleus,
compared to $\rho_p$, the relative slow change of $\rho_n$ in
the surface forms the neutron-skin structure. It should be noted that a neutron-rich nucleus
does not guaranty it has a neutron skin since more neutrons are needed to compensate the Coulomb
interaction in the large-$Z$ nucleus. It can be assumed that the neutron-skin structure should be more
appropriate to explain the phenomena shown in the IYR(m), i.e., the linear part of IYR(m) can be explained
as the little variation of $\rho_n$ and $\rho_p$ in the core, and the different trends of the
nonlinear part in IYR(m) could be explained as the neutron-skin effect when considering the
difference between $\rho_n$ and $\rho_p$ in the nuclear surface.

\begin{figure}[htbp]
\includegraphics
[width=8.6cm]{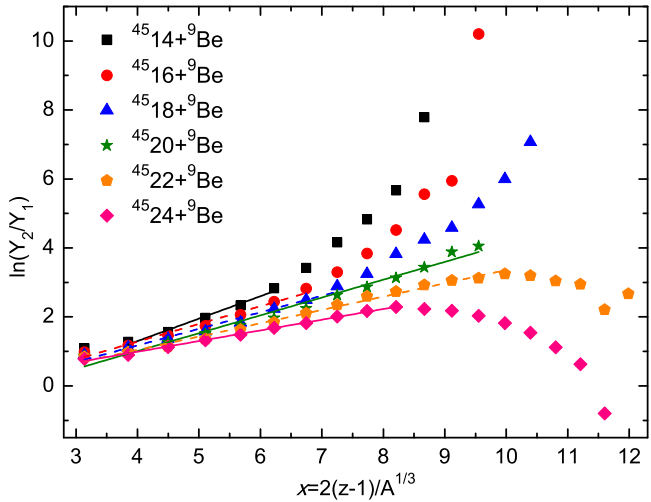}
\caption{\label{A45IYRm} (Color online)
The IYR(m) in the calculated 140$A$ MeV $^{45}$Z + $^9$Be reactions.
The projectile changes from $Z =$ 14 to 24 in a step of 2.
The lines are the linear fitting results of IYR(m).
}
\end{figure}

To illustrate whether IYR(m) depends on the neutron skin of the projectile, the reactions
of some $A_p = 45$ isobars, i.e., the 140$A$ MeV $^{45}Z$ + $^9$Be reactions, are calculated.
$Z$ of the selected projectiles are $Z =$ 14, 16, 18, 20, 22, and 24, of which $I'$ changes from
0.38 to --0.07. The $\delta_{np}$ of the projectiles are 0.179, 0.127, 0.073, 0.018, --0.038, and --0.096fm,
respectively. The IYR(m) in these reactions are plotted in Fig. \ref{A45IYRm}, which has the
similar distribution as in the calcium isotopic reactions. Since the volume of the isobaric
projectiles shall be the same, the volume dependence of IYR(m) can be eliminated in
these reactions. The difference among the nonlinear part of IYR(m) should be explained as
the neutron-skin effects.

\begin{figure}[htbp]
\includegraphics
[width=8.6cm]{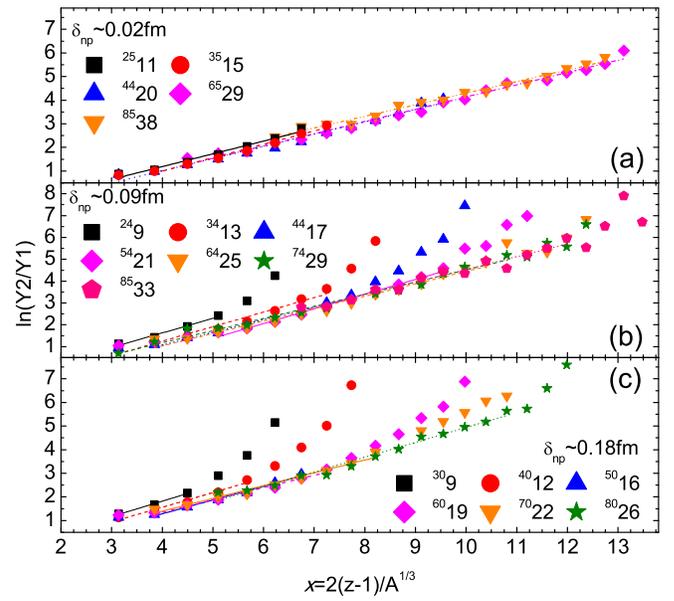}
\caption{\label{skinIYRm} (Color online)
The IYR(m) in the calculated 140$A$ MeV $^{A}$Z + $^9$Be reactions induced
by projectiles having similar neutron-skin thickness ($\delta_{np}$). $A$ and
$Z$ are the charge and mass numbers of the projectile. In (a), (b) and (c),
the reactions are for projectiles having $\delta_{np} \approx 0.02, 0.09$,
and $0.18$fm, respectively. The lines represent the linear fitting results
of the IYR(m).
}
\end{figure}

Furthermore, to show how the neutron-skin affects the IYR(m), the reactions induced by the
projectiles having similar $\delta_{np}$ ($\sim0.02, 0.06, 0.09, 0.13$, and $0.18$fm) are calculated.
The selected $\delta_{np}\sim0.02$fm projectiles are $^{25}$11, $^{35}$15, $^{45}$20, $^{65}$29,
and $^{85}$38; the $\delta_{np}\sim0.06$fm projectiles are $^{32}$14, $^{42}$17, $^{52}$21, $^{62}$25,
and $^{70}$30; the $\delta_{np}\sim0.09$fm projectiles are $^{24}$9, $^{34}$13, $^{44}$17, $^{54}$21,
$^{64}$25 and $^{74}$29; the $\delta_{np}\sim0.13$fm projectiles are $^{26}$9, $^{37}$13, $^{48}$17,
$^{66}$24, $^{77}$28 and $^{86}$31; the $\delta_{np}\sim0.18$fm projectiles are $^{30}$9, $^{40}$12,
$^{50}$16, $^{60}$19, $^{70}$22 and $^{80}$26, which covers a large range of mass from 25 to 86, and
$I'$ from 0.1 to 0.4. The target nucleus is $^{9}$Be and the incident energy is 140$A$ MeV.
In Fig. \ref{skinIYRm}, the IYR(m) in reactions of the $\delta_{np}\sim0.02, 0.09$ and $0.18$fm
projectiles are plotted. In Fig. \ref{skinIYRm}(a), the IYR(m) in the $\delta_{np}\sim0.02$fm
projectile reactions overlap, and a quite good linear correlation between IYR(m) and $x$ is shown
in each reaction. No isospin and volume dependence of IYR(m) is shown.
From Figs. \ref{skinIYRm}(b) to \ref{skinIYRm}(c), when the projectile becomes more neutron-rich, the IYR(m)
shows a quick increase in fragments having large $x$, being the same as those shown in the calcium
isotopes and the $^{45}$Z-induced reactions. In these neutron-rich-projectile-induced reactions,
the IYR(m) is greatly influenced by $A_p$ when $A_p$ is relatively small. When $A_p$ is relative
large, the IYR(m) overlap in a large range of $x$ (for example, $A_p > 34$ when $\delta_{np}\sim0.09$fm,
and $A_p > 50$ when $\delta_{np}\sim0.18$fm). It can be concluded that the IYR(m) depend on $A_p$
very little in the neutron-rich projectile induced reactions when $A_p$ is large, which can be
explained as that these projectiles have relative large cores in which $\rho_n$ and $\rho_p$
change very little. The isospin dependence of the IYR(m) in these reactions will be discussed
later.

\begin{figure}[htbp]
\includegraphics
[width=8.6cm]{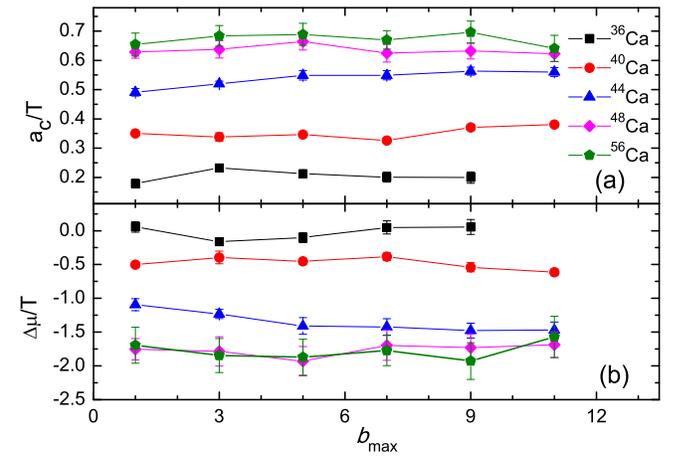}
\caption{\label{LFCa} (Color online)
The $a_c/T$ and $\Delta\mu/T$ from the fitting result of the linear part of the IYR(m)
in the calcium isotopic reactions which are plotted in Fig. \ref{Ca3656bmax}. The $x$
axis represents the limitation of the maximum impact parameter ($b_{\mbox{max}}$).
}
\end{figure}

To show the neutron-skin effects in the IYR(m) more clearly, the linear part of the IYR(m)
is fitted using a linear function. In Figs. \ref{LFCa}(a) and \ref{LFCa}(b), the $a_c/T$ and $\Delta\mu/T$
determined from the calcium isotopic reactions are plotted, respectively. The $a_c/T$
($\Delta\mu/T$) is found to increase (decrease) when the projectile becomes more neutron-rich,
but $a_c/T$ and $\Delta\mu/T$ varies very slowly as $b_{\mbox{max}}$ becomes larger. It can be
concluded that in central collisions where the skin has less influence on the yields of
fragments, the IYR(m) does not depend on the colliding volume, whether the projectile is
neutron-rich or not.

\begin{figure}[htbp]
\includegraphics
[width=8.6cm]{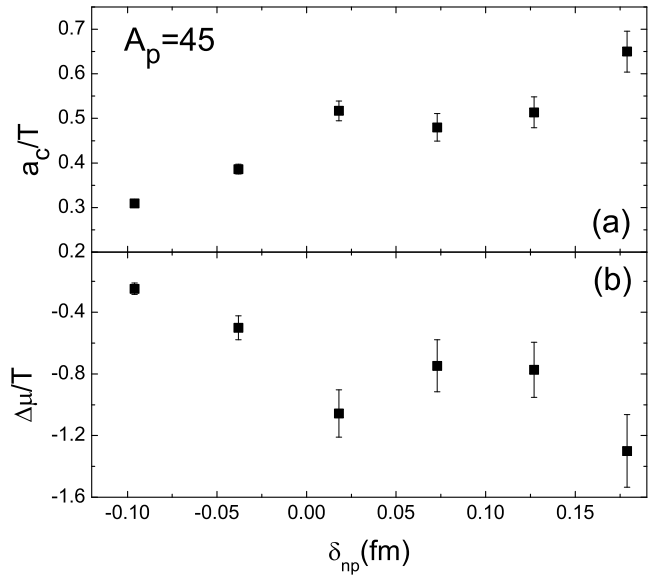}
\caption{\label{LFA45} (Color online)
The $a_c/T$ and $\Delta\mu/T$ from the fitting result of the linear part of IYR(m) in the
isobaric reactions plotted in Fig. \ref{A45IYRm}.  The $x$ axis represents the
neutron-skin thickness of $A_{p} = 45$ isobars.
}
\end{figure}

\begin{figure}[htbp]
\includegraphics
[width=8.6cm]{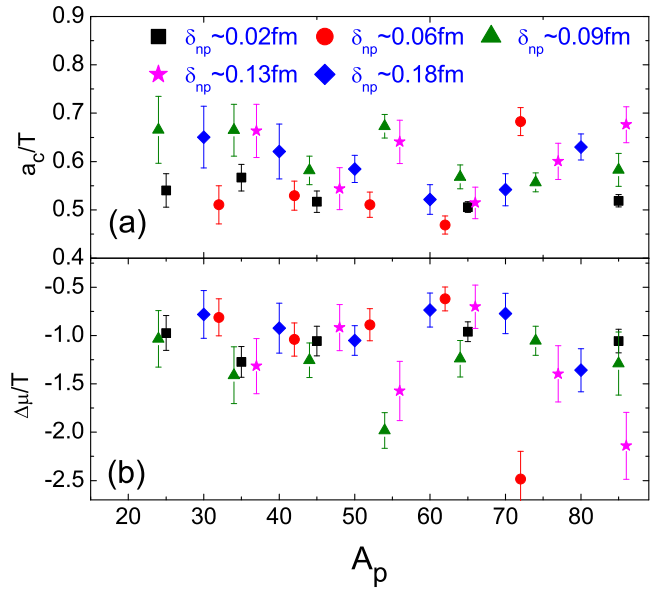}
\caption{\label{LFskin} 
$a_c/T$ and $\Delta\mu/T$ from the fitting result of the linear part of the IYR(m)
in the reactions induced by projectiles having $\delta_{np}\sim0.02, 0.06, 0.09,
0.13$, and $0.18fm$. Some IYR(m) of these reactions are plotted in Fig. \ref{skinIYRm}.
The $x$ axis represents the mass of the projectile ($A_p$).
}
\end{figure}

The $a_c/T$ and $\Delta\mu/T$ determined from the IYR(m) in the isobaric-projectiles-induced reactions
are plotted as a function of $\delta_{np}$ in Figs. \ref{LFA45}(a) and \ref{LFA45}(b), respectively.
The $a_c/T$ increases as $\delta_{np}$ of the projectile becomes larger, but $a_c/T$ only shows
little difference when $\delta_{np}>0$. The $a_c/T$ and $\Delta\mu/T$ determined from the
IYR(m) in the reactions of the similar $\delta_{np}$ projectiles are plotted in Figs. \ref{LFskin}(a) and \ref{LFskin}(b), respectively.
Both the values of $a_c/T$ ($\Delta\mu/T$) are similar (especially when $A_p > 40$). Based on
the results in Figs. \ref{LFA45} and \ref{LFskin}, it can be concluded that, in the reaction of a
projectile having $\delta_{np} > 0$, if the IYR(m) of fragments vulnerable to the neutron-skin
effect is not considered, the volume dependence of the IYR(m) disappears when $A_p$ is relatively
large. In the reaction induced by a projectile having similar $\delta_{np}$, the IYR(m) does not
depend on the colliding volumes in the central collisions. Thus in these reactions, the $a_c/T$ and
$\Delta\mu/T$ determined are not influenced by the mass or volume of the projectile. In particular,
in the $\delta_{np}\sim0$ projectiles, in which the neutrons and protons distributions are almost
equal, the IYR(m), and the resultant $a_c/T$ and $\Delta\mu/T$ are not affected by the mass or volume of
the projectile.

\begin{figure}[htbp]
\includegraphics
[width=8.6cm]{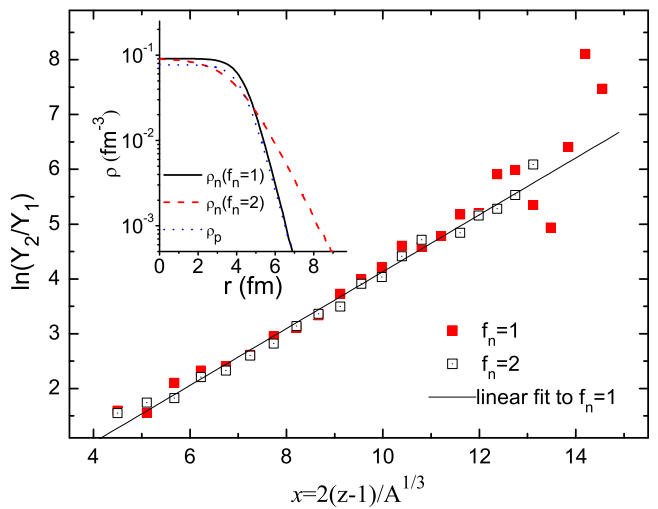}
\caption{\label{changeskin} (Color online)
The IYR(m) in the calculated 140$A$ MeV $^{65}$29 + $^9$Be reactions. $\delta_{np}$ of $^{65}$29
is adjusted by changing $f_n$ in Eq. (\ref{Fermi}). The open and solid squares represent the
results of $f_n =$ 1 and 2, respectively. The line is the linear fitting result to the IYR(m)
of $f_n =$ 1. In the inserted figure, the lines represent $\rho_n$ according to $f_n =$
1 (solid), 2 (dashed), and $\rho_p$ (dotted) of $^{65}$29 according to Eq. (\ref{Fermi}).
}
\end{figure}

It is interesting to note how IYR(m) change if the neutron-skin thickness is adjusted. By
changing the value of $f_i$ in Eq. (\ref{Fermi}), the diffuseness in $\rho_n$, at the same time
$\delta_{np}$ can be changed. For the $^{A}Z=^{65}29$ nucleus, $\delta_{np} =$0.02 and
0.85 fm when $f_n=1$ and 2, respectively. In Fig. \ref{changeskin}, the IYR(m) according to $f_n=1$ and 2
in the reactions are plotted, and the $\rho_n$ and $\rho_p$ distributions are plotted in
the inserted figure. Compared to the result of $f_n = 1$, more fragments of large mass are produced in the
result of $f_n =$ 2. The IYR(m) in the $f_n =$ 1 and 2 results almost overlap, but an increase
can be found in the isobars of $x > 10$ [the IYR(m) of $x > 12$ show the accelerating increase
trend]. It can be concluded that the change of density distribution does change the IYR(m)
distribution.

\begin{figure}[htbp]
\includegraphics
[width=8.6cm]{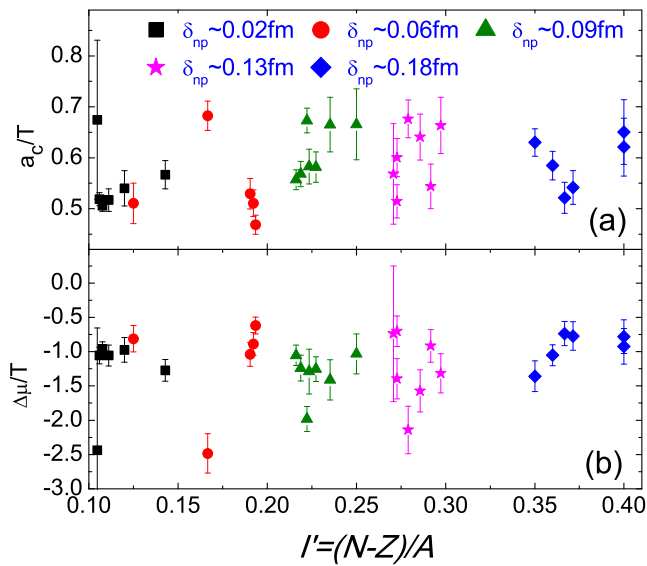}
\caption{\label{LFisospin}  (Color online)
The $a_c/T$ and $\Delta\mu/T$ from the fitting of the linear part of IYR(m) in
the reactions of projectiles with $\delta_{np}\sim0.02, 0.06, 0.09, 0.13$, and
$0.18$ fm. The $x$ axis represents the isospin parameter [$I'=(N-Z)/A$] of the
projectile.
}
\end{figure}

Finally, we discuss the isospin dependence of the IYR(m). The nonlinear part of IYR(m) is
omitted in the determination of the $a_c/T$ and $\Delta\mu/T$. It was noted previously that
the nonlinear part was due to the skin effects in the IYR(m). An equivalent "isospin" can
be defined as $(\rho_n - \rho_p)/(\rho_n + \rho_p)$. For neutron-rich (or proton-rich) projectiles, in some sense,
the "isospin" changes quickly in the surface region due to the fast change of $\rho_n$ and
$\rho_p$; while the "isospin" keeps constant in the core region. In Figs. \ref{LFisospin}(a)
and \ref{LFisospin}(b), the $a_c/T$ and $\Delta\mu/T$ in Fig. \ref{LFskin} are re-plotted as a function
of the isospin of the projectile. From $I'=0.1$ to 0.4, the $a_c/T$ shows a small difference,
and even less variation in $\Delta\mu/T$ is found. It is concluded that if $a_c/T$ and $\Delta\mu/T$
determined from the linear part of IYR(m), the isospin effects are also very small. Thus
the neutron-skin effects can well explain the systematic phenomena shown in IYR(m).

\section{summary}
\label{summary}
In summary, we focus on the interpretation of the systematic dependence of IYR(m) and
the extracted parameters from them. The previously proposed isospin or volume dependence
of IYR(m) did not fully explain the phenomena shown in the experimental results, or the theoretical
results of the SAA and the SGC/CSE models. Considering the density effects in the fragment
production, it is assumed that $\delta_{np}$ of the projectile affects the IYR(m). Using the SAA
model, which considers the density difference of a proton and neutron conveniently, the reactions
induced by three series of projectiles are calculated: (1) the calcium isotopes from $^{36}$Ca
to $^{56}$Ca with different limitations of $b_{\mbox{max}}$; (2) the $A=45$ isobaric projectile of
which $Z$ varies from 14 to 24; and (3) the projectiles having similar $\delta_{np}$.
Generally, the IYR(m) in the calculated reaction shows the distribution of a linear part in the
small-$x$ fragments and a nonlinear part in the large-$x$ fragments except in the reactions of
$\delta_{np}\sim0$ projectiles. The linear part of IYR(m) is explained as the core effects of
the projectile, and the nonlinear part of IYR(m) is assumed to be governed by the skin of the
projectile. In the calcium isotopes induced reactions, it is shown that IYR(m) does not
depend on the volume of the colliding source in the central collisions, whether the projectile
is neutron-rich or not. In the isobaric reactions, the IYR(m) is found to be greatly influenced by
$\delta_{np}$ of the projectile, but when the projectile has a relatively large $\delta_{np}$, the
IYR(m) depends on the isospin very little. If the $\delta_{np}$ of the projectile is similar,
IYR(m) does not depend on the mass or volume of the projectile when its mass is relatively
large. From the calculated results, it can be concluded that when $\rho_n$ and $\rho_p$ change
very little in the core, or $\rho_n$ and $\rho_p$ in the entire projectile are similar (such as
the $\delta_{np}\sim0.02$fm projectiles in which $\rho_n$ and $\rho_p$ can be assumed to be
the same), both the isospin and volume dependence of IYR(m) disappears. It can be concluded
that the system dependence of IYR(m) shown in the SAA and experimental data shall be the
neutron-skin effects, and neither the isospin nor the volume dependence of IYR(m) can completely
explain this dependence. The finite effects suggested in the SGC/CSE results are inadequate to
explain the experimental results, and also disagree with the SAA results.

At last, we comment on the $a_c/T$ and $\Delta\mu/T$ in the isotopic reactions, which show a
dependence on the mass of the projectiles. Due to the values of $a_c$, $\Delta\mu$ and $T$
are difficult to separate in the free-energy models; they influence each other in the
fitting. $\Delta\mu$ reflects the properties of the projectile, which can be assumed to
increase when the projectile becomes more neutron-rich. In a canonical thermodynamic (CTM)
model, a temperature profile of impact parameter ($b$), which decreases quickly as $b$ increases,
is introduced to improve the prediction of the fragment yield \cite{Mallik11PRC-PF-Tb}. Due the
low $\rho_n$ and $\rho_p$ in the surface (which corresponds to the peripheral collisions),
the abraded nucleons are also less than those in the central collisions, which can also result
in the relatively low temperature \cite{MaCW13isoSB,MaCW12PRCT,Mallik11PRC-PF-Tb}. In this sense,
the low temperature in the peripheral collisions is one result of the low density \cite{MaCW13isoSB,MaCW12PRCT}.
The increasing $a_c/T$ in the large-$x$ fragments can also be explained as the density effects
since the surface regions of the projectile (target) govern the peripheral collisions. Due to
the complexity of the temperature dependence on the impact parameters and densities, and the
possible dependence of $a_c$ on the density \cite{Huang-PRC11-freeenergy,PMar12PRCIsob-sym-isos},
we can not know exactly what the actual value of $a_c$ is. It was proposed that the RCI between
the isobars can be determined by some approximations as in Refs. \cite{Huang10,PMar12PRCIsob-sym-isos,ResC13},
and it was shown that, though the Coulomb term was retained in determining the symmetry-energy
coefficient of the neutron-rich fragment in the IYR method \cite{Huang10,MaCW12EPJA,MaCW12CPL06,MaCW13CPC},
the RCI between the isobars in the difference of IYRs was negligible \cite{PMar12PRCIsob-sym-isos,ResC13}.

\begin{acknowledgments}
This work is supported by the National Natural Science Foundation of China (Grants No. 10905017 and 11035009),
the Knowledge Innovation Project of the Chinese Academy of Sciences under Grant No. KJCX2-EW-N01, the Program for
Science \& Technology Innovation Talents in Universities of Henan Province (13HASTIT046), and the Young Teacher Project
in Henan Normal University.
\end{acknowledgments}

\end{document}